*Di ZHU*
*Ewan PRITCHARD*
*Sumanth Reddy DADAM*
*Vivek KUMAR*
*Yang XU*

# Optimization of rule-based energy management strategies for hybrid vehicles using dynamic programming

*Reducing energy consumption is a key focus for hybrid electric vehicle (HEV) development. The popular vehicle dynamic model used in many energy management optimization studies does not capture the vehicle dynamics that the in-vehicle measurement system does. However, feedback from the measurement system is what the vehicle controller actually uses to manage energy consumption. Therefore, the optimization solely using the model does not represent what the vehicle controller sees in the vehicle. This paper reports the utility factor-weighted energy consumption using a rule-based strategy under a real-world representative drive cycle. In addition, the vehicle test data was used to perform the optimization approach. By comparing results from both rule-based and optimization-based strategies, the areas for further improving rule-based strategy are discussed. Furthermore, recent development of OBD raises a concern about the increase of energy consumption. This paper investigates the energy consumption increase with extensive OBD usage.*

## 1. Introduction

As a key tool in the race to continuously reduce energy consumption in transportation, electrification has emerged as the primary means to achieve this goal. Hybrid electric vehicles (HEVs) play a significant role in the family of electrified vehicles. Unlike conventional vehicles and battery electric vehicles (BEVs), hybrid electric vehicles have a greater complexity and take advantage of operating dual propulsion systems to meet the load on the road. The dual propulsion system usually consists of a high voltage (HV) battery pack, a traction motor, an engine, and a generator. The combination of the HV battery pack and the motor provides customers with pure electric vehicle (EV) driving experience. When the state of charge (SOC) falls below a certain threshold, the engine will engage to charge the HV battery pack and/or provide additional power to meet the instantaneous load. Because HEVs can allow some or all the powertrain components to operate at a time, the operation is more complex than conventional vehicles and BEVs. The control strategy that manages these components to achieve the lowest energy consumption is the energy management strategy.

The energy management strategy can be divided into three categories: rule-based strategies, optimization-based strategies, and a mix of the two. The rule-based strategy is often also called an online strategy and employs control laws and rules to achieve a local optimal point [1, 11, 14, 15, 17, 19, 21]. One advantage of rule-based strategies is that they require less computational power to make decisions. Another advantage is that it does not require future information about the trip. A group of researchers used ADVISOR software to study several rule-based strategies [1]. They suggested the charge depleting (CD) and charge sustaining (CS) strategies along with electric assist strategy are more effective. Another group of researchers built a co-simulation platform using CANoe and MATLAB/Simulink to study a rule-based strategy which allows the vehicle to switch between CD mode and CS mode [14]. Generally, once entering CS mode, a plug-in hybrid electric vehicle (PHEV) does not switch back to CD mode until it is charged via alternative current (AC). A blended rule-based strategy was proposed in [11]. This strategy does not consider a specific vehicle speed or acceleration but rather vehicle energy. Comparing to a conventional rule-based strategy, the authors claimed that the fuel economy was improved by 18.4%.

The optimization-based strategies such as nonlinear programming, genetic algorithm, and dynamic programming (DP) require prior knowledge of the trip [9, 12, 15, 20]. In addition, they are more computationally intensive than rule-based strategies. Usually, they are used offline to find the global optimal point that could have been achieved after the fact.

Mixed strategies such as the equivalent consumption minimization strategy (ECMS) take advantage of both rule-based strategies and optimization-based strategies [5, 6, 10, 13]. The operating parameters are determined by using optimization techniques. After this determination, these pre-determined parameters are used by rule-based algorithms in real-time controllers within the vehicle. Therefore, the computational burden in the vehicle is avoided.

Recent research on the optimization-based strategies has focused on utilizing knowledge from modeling and simulation to make the optimization results more representative. Two dynamic programming algorithms were studied in [20]. The deterministic dynamic algorithm solves the optimization problem by sequentially calculating every state at every time step in a backwards order. Unlike the deterministic dynamic algorithm, the stochastic dynamic programming algorithm approach results in a control which depends on a specific state [20]. A stochastic model of a vehicle's drive missions was used in their study and the strategy was validated in hardware-in-the-loop (HIL). Another group of researchers combined DP with a neural network to perform the energy management [9]. Their method utilizes the neural network to predict the decision variable which normally is calculated by DP. A rule-based strategy was proposed based on the projection partition of composite power system efficiency in [12]. In conjunction to the rule-based strategy, a DP method was proposed on the basis of the establishment of the whole system. Some other researchers proposed a recalibration method to improve a rule-based strategy in [15]. The DP essentially helps calibrate the rule-based strategy. However, none of the above references recognizes that the measurement errors in the vehicle plays an important role in the energy management strategy. In addition, the complexity of vehicle dynamics results in a much higher energy consumption at the wheel than their models predicted. Therefore, their optimization results may not represent the energy consumption in real world.

A recent development in on-board diagnostics (OBD) suggests that more OBD services and routines may be required for future vehicles [3]. The use of deceleration fuel shutoff in conventional vehicles to diagnose powertrain faults and meet OBD regulatory requirements is replaced by spinning the gen-set in the HEV [3, 8]. The increasing number of OBD events may lead to an increase in energy consumption for HEVs. Therefore, it is important to understand if extensive OBD events have an impact on the energy consumption.

To achieve the above objectives, we propose a framework that utilizes physical test data from a series PHEV under the emission and energy consumption (E&EC) drive cycle to get more representative DP result. The E&EC drive cycle represents real world driving in the US. After that, the DP result is analyzed to identify where the rule-based strategy can be improved. At last, the impact of recent OBD development on the energy consumption is assessed.

The remainder of this paper is organized as follows. The Method section is divided into five subsections. We first review the E&EC drive cycle test. After that, we discuss the vehicle architecture and specifications. Following that powertrain component models are discussed. Later, we visit both the rule-based strategy and the optimization-based strategy used in this study. Most importantly, our discussion sheds light on how the test data from the rule-based strategy is used to improve the optimization-based strategy. We also discuss how the optimization via dynamic programming is fed back to improve the rule-based strategy. In the last subsection, we talk about the method used to calculate energy consumption. In the Results and Discussion section, we begin with discussing the result from the drive cycle analysis. Following that, we review the energy consumption from both the vehicle test and simulations. Finally, we conclude this paper in the Conclusions section.

## 2. Method

The proposed framework has four steps. The first step is to analyze the drive cycle(s). The result from the analysis include propulsion energy, peak power, average positive power, and percent idle time. These four items are selected to represent the characteristics of the drive cycle(s). The second step is to perform a vehicle test in the same drive cycle(s). Vehicle test mass, vehicle velocity, power input and output from each powertrain component, and other properties need to be collected during the test. The third step is to compare the drive cycle result with the test result. The difference between the two is used to feed into DP to generate a more representative optimization result. At last, the optimization result is looped back with the vehicle test result to identify areas in the energy management strategy that can be improved.

### 2.1. E&EC test

The E&EC drive cycle is illustrated in Fig. 1. The E&EC drive cycle was developed by Argonne National Lab to address more real-world driving conditions on the road. The drive cycle blends four standard certification test cycles: Federal Test Procedure (FTP), Highway Fuel Economy Driving Schedule (HWFET), US06 City and US06 Highway. This drive cycle aims to have a much higher top speed with more aggressive accelerations and decelerations. The distance of the drive cycle is 22.55 km.

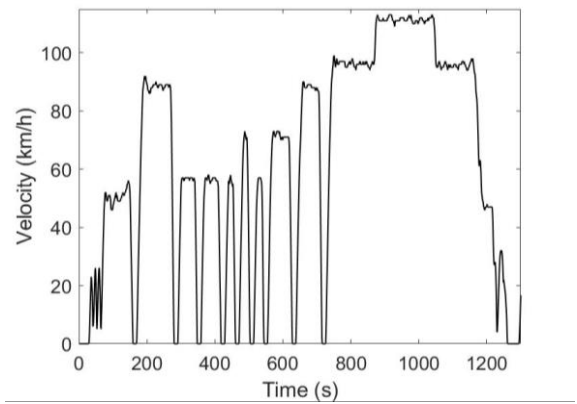

Fig. 1. E&EC drive cycle

The vehicle had one driver and one passenger during the test. The vehicle also towed a trailer in which a SEMTECH emissions analyzer was used to measure the exhaust gases from the exhaust pipe. As a result, the driver, passenger, and trailer added additional 700 kg to the vehicle in the test, which increases the energy consumption of the vehicle.

### 2.2. Vehicle Overview

A 2013 General Motors Malibu was used in this study as shown in Fig. 2. After the original powertrain of the vehicle was removed from the vehicle, a set of new powertrain components was selected and integrated into the vehicle to represent a series PHEV. A peak 100 kW Magna electric motor is coupled to a single speed transmission with a gear ratio 7.82:1 to power the front axle. The gen-set consists of a continuous 37 kW TM4 generator and a 33 kW Kubota diesel engine. The generator and engine are mechanically coupled through a herringbone belt drive at a ratio of 2.7:1. An 18.9 kWh HV battery pack using lithium-ion phosphate cells is electrically coupled with the motor, generator and a 3.3 kW BRUSA on-board charger in an HV junction box. The high-level vehicle architecture is illustrated in Fig. 3. This architecture advances other architectures by allowing the range extender to be completely decoupled from the wheels and run at optimal operation points. This vehicle may also be considered as an extended-range electric vehicle (EREV) that is a subcategory of HEVs. We use PHEV to emphasize the plug-in feature.

*Optimization of rule-based energy management strategies for hybrid vehicles using dynamic programming*

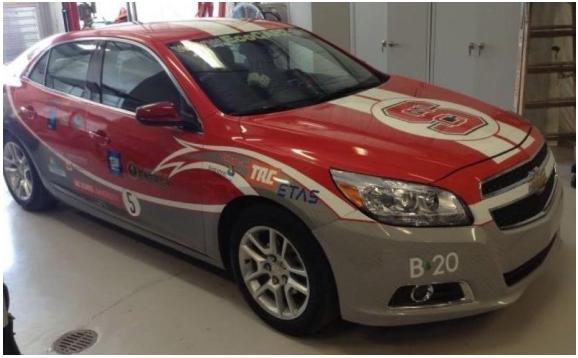

Fig. 2. The reengineered vehicle

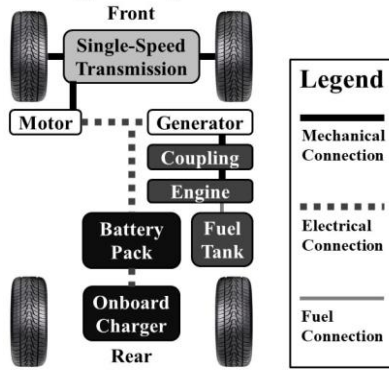

Fig. 3. Vehicle architecture

The large HV battery pack allows the vehicle to drive approximately 80 km before activating the gen-set. The total range of the vehicle is about 378 km. The mass of the vehicle is 2100 kg. The electric motor is the only powertrain component coupled with the wheels. The 0-to-100 kph acceleration is 13.2 seconds.

**2.3. Vehicle model**

The vehicle model consists of a vehicle dynamic model, motor model, gen-set model, and a battery model. The vehicle dynamic model was used to calculate the power and energy needed at the wheels to propel the vehicle. The model is described by the equation below:

$$m_i \cdot \frac{dv}{dt} = F_{tr} - F_{grade} - F_{aero} - F_{rr} \quad (1)$$

$$m_i = m \cdot i \quad (2)$$

where $m_i$ is the inertial mass of the vehicle in kg, $v$ is the vehicle velocity in m/s, $F_{tr}$ is the tractive force at the wheels in N, $F_{grade}$ is the force generated by a grade in N, $F_{aero}$ is aerodynamic drag in N, $F_{rr}$ is the rolling resistance in N, m is the vehicle mass in kg, and i is the rotating inertia factor. Substituting vehicle properties into Equation (1), the equation becomes

$$m_i \cdot \frac{dv}{dt} = F_{tr} - m \cdot g \cdot \sin\alpha - \frac{1}{2} \cdot \rho \cdot C_d A_f \cdot v^2 - m \cdot g \cdot C_{rr} \quad (3)$$

where g is the gravitational constant in N/kg, $\alpha$ is the road grade in degrees, $\rho$ is air density in kg/m$^3$, $C_d A_f$ is the product of air drag coefficient and vehicle frontal area in m$^2$, and $C_{rr}$ is the rolling resistance coefficient. The vehicle properties used in this study can be found in Table 1.

Table 1. Vehicle properties

| Parameter | Symbol | Value |
|---|---|---|
| Vehicle mass | m | 2100 kg |
| Vehicle test mass | $m_t$ | 2800 kg |
| Drag coefficient × frontal area | $C_d A_f$ | 0.75 m$^2$ |
| Rolling resistance coefficient | $C_{rr}$ | 0.009 |

The motor and gen-set were modeled using the black box modeling technique [22]. For the motor model, the inputs include accelerator pedal position, brake pedal position, and vehicle velocity. The outputs of the motor model are motor torque, motor speed, motor input power, motor output power, and motor losses. An efficiency map shown in Fig. 4 was developed for this model using component characterization test data such as motor voltage input, motor current input, motor torque output, and motor speed output. The following equation was used to calculate the efficiencies:

$$\eta_{motor} = \frac{T_{motor} \cdot \omega_{motor}}{V_{motor} \cdot I_{motor}} \cdot 100\% \quad (4)$$

where $\eta_{motor}$ is the motor efficiency in %, $T_{motor}$ is the torque output in Nm, and $\omega_{motor}$ is the speed output in rpm, $V_{motor}$ is the voltage input in V, and $I_{motor}$ is the current input in A.

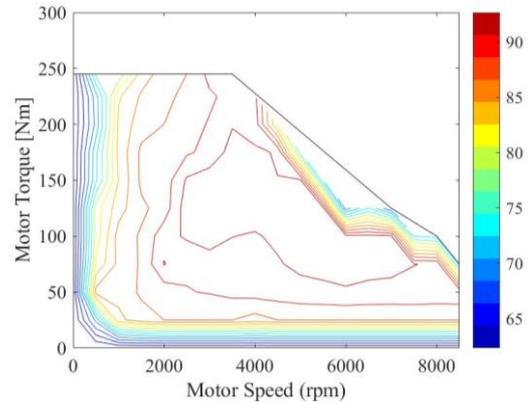

Fig. 4. Motor efficiency map

The gen-set model is similar to the motor model. The following two equations were used to calculate the engine and generator efficiencies, respectively:

$$\eta_{eng} = \frac{1}{BSFC \cdot LHV} \cdot 100\% \quad (5)$$

$$\eta_{gen} = \frac{V_{gen} \cdot I_{gen}}{T_{gen} \cdot \omega_{gen}} \cdot 100\% \quad (6)$$

where $\eta_{eng}$ is the engine efficiency in %, BSFC is the brake-specific fuel consumption in g/(kWh) and LHV is the low heating value in kWh/g, $\eta_{gen}$ is the generator efficiency in %, $V_{gen}$ is the voltage output of the generator in V, $I_{gen}$ is the current output of the generator in A, $T_{gen}$ is the generator torque input in Nm, and $\omega_{gen}$ is the speed input in rpm. The engine and generator efficiency maps are illustrated in

COMBUSTION ENGINES, 2021, 184(1)   5

Figs. 5 and 6. The efficiency map of the gen-set was created by merging the generator and engine efficiency maps together in Fig. 7.

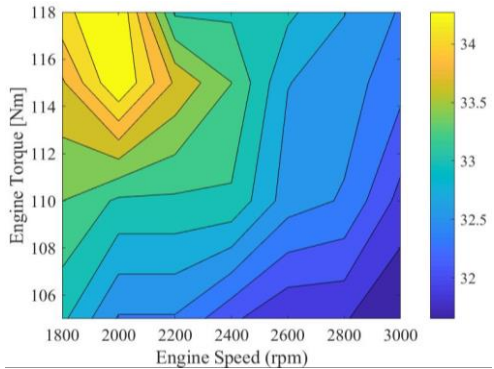

Fig. 5. Engine efficiency map

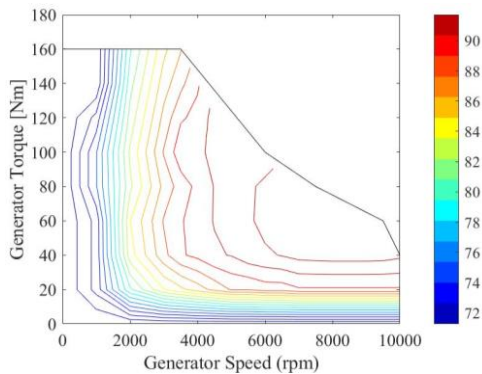

Fig. 6. Generator efficiency map

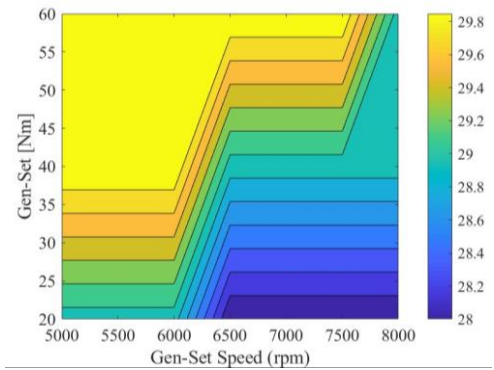

Fig. 7. Gen-set efficiency map

The battery model was represented by a simple internal resistance model. The model can calculate cell open-circuit voltage, battery pack open-circuit voltage, cell voltage, pack voltage, battery losses, and SOC. The model is described below:

$$P_{batt} = \frac{R_{in} \cdot I_{batt}^2 + V_{oc} \cdot I_{batt}}{1000} \quad (7)$$

$$SOC = \frac{\int V_{oc} \cdot I_{batt}}{3600000 \cdot C_{batt}} \cdot 100\% \quad (8)$$

where $P_{batt}$ is battery pack power in kW, $R_{in}$ represents the internal resistance in $\Omega$, $I_{batt}$ is the battery pack current in A, $V_{oc}$ is the pack open-circuit voltage in V, SOC is the state of charge in %, and $C_{batt}$ is the battery pack capacity in kWh.

### 2.4. Energy management strategies

A rule-based energy management strategy was implemented in the vehicle. The rule-based strategy has two primary operation modes: CD mode and CS mode. The CD mode was designed to take advantage of the 18.9 kWh battery pack to achieve 80 km pure electric range. Customers get the smoothest and quietest driving experience while recovering kinetic energy through advanced regenerative braking strategies. The 18.9 kWh battery pack was selected to ensure the pure electric range can satisfy 80% of consumers daily travel needs. This 80% design criterion is the result of the utility factor calculated from SAE J2841 in [18]. In this phase, the propulsion is solely supported by the motor and battery pack. Therefore, the SOC continues depleting as the vehicle travels.

The CS mode does not begin until the SOC falls below a designed threshold in the SOC CS window. After that, the gen-set is turned on to increase the battery pack SOC until the upper limit of the SOC CS window is reached. $SOC_{high}$ and $SOC_{low}$ are used to represent the upper and lower limits, respectively. They form the SOC window. The threshold to initiate CS mode is set slightly higher than $SOC_{low}$ to ensure the gen-set has time to begin power generation and keep the SOC within the window while operating at optimal points. The SOC moves within the SOC window in the CS mode. One of the key advantages of this architecture is that the gen-set can be operated at any given operating point regardless of road load. To avoid turning on and off the gen-set too frequently, a 10-second ramp down time was implemented. Once the gen-set is turned on, the control strategy does not turn it off in the next 10 seconds and vice versa. Besides the gen-set ramp down, two other techniques were also implemented to provide the customer with a smoother experience. One of the two techniques is the delay torque production from the engine. After cranking the engine, the generator spins freely with the engine until the engine warms up. The other technique is the maximum current limit for regenerative braking. While the gen-set is pushing current into the battery pack, the regenerative braking algorithm continuously monitors the current on to HV bus to avoid putting the battery pack into an overcurrent condition.

The data from the vehicle test opens up an opportunity to further improve the rule-based energy management strategy through optimization. Since the vehicle depletes the battery pack to the lower limit in CD mode no matter what the driving conditions are, there is no gain to optimize the energy management strategy for the CD mode. However, if the trip information such as vehicle velocity can be either predicted or known prior to beginning the trip, optimizing the energy management strategy in CS mode can provide a noticeable energy consumption reduction. In addition, recent studies have shown interest of using the generator to perform OBD. Still, it is unclear that how the new development in OBD is going to affect the energy consumption for hybrids. Thus, an optimization-based energy management strategy considering the latest OBD techniques is proposed in the following paragraph.

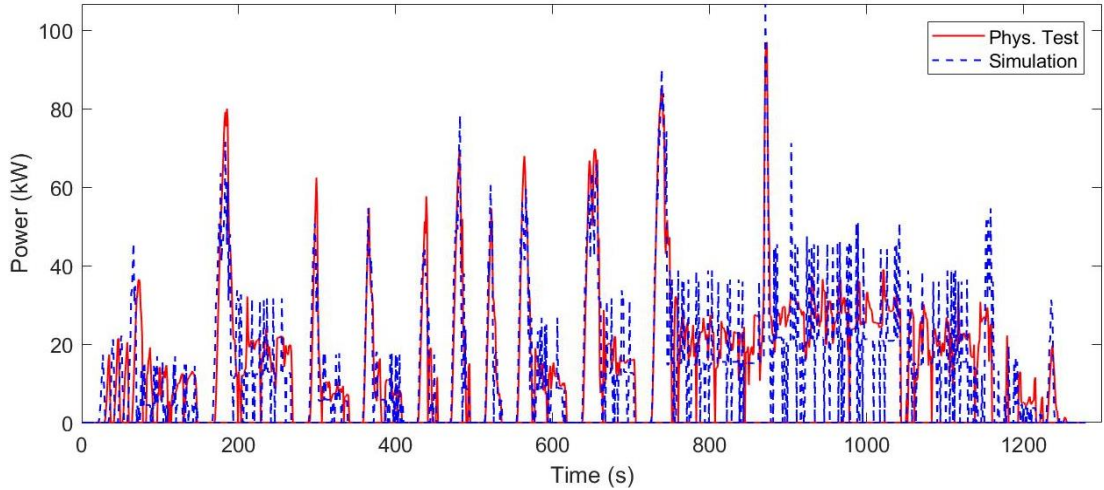

Fig. 8. Positive power from the simulation and physical test for a single E&EC drive cycle

The state equations in the discrete-time format can be expressed as:

$$x_{k+1} = f(x_k, u_k) \quad (9)$$

$$x_k = SOC_k \quad (10)$$

$$u_k = \Delta SOC_k \quad (11)$$

where x denotes the state variable, k is the step in discrete-time format, and u represents the decision variable. The cost function for minimizing the energy consumption in the general format is expressed as:

$$J_k(x_k) = \min_{u_k}\{f(x_k, u_k) + J_{k+1}(x_{k+1})\} \quad (12)$$

The constraints that should be met are as follows:

$$\begin{cases} SOC_{min} \leq SOC_k \leq SOC_{max} \\ 0 \leq \Delta SOC_k \leq \frac{P_{gen-set\_max}}{3600 \cdot C_{batt}} \cdot 100\% \end{cases} \quad (13)$$

When the rule-based strategy was first developed and implemented, there was not enough information about the maturity of the prototype vehicle. Therefore, the operating point from the gen-set was selected conservatively. After testing the vehicle on the proving ground, all the operating points in the gen-set efficiency map are available. However, we still want to select operating points that are at the same gen-set speed as we originally selected for the rule-based strategy. Therefore, the following equations are used to make decisions in the optimization:

$$u_k = \begin{cases} a & SOC_{max} \leq SOC_k + 0.567 \\ b & SOC_{min} \leq SOC_k < SOC_{max} - 0.567 \end{cases} \quad (14)$$

$$a = 0 \quad (15)$$

$$b = (0.294:0.051:0.567) \quad (16)$$

where a and b represent SOC change in % for 10-second period.

In summary, two optimization cases were studied. The first case optimized the energy consumption for the CS portion of 3-lapse E&EC drive cycles. The travel distance for that portion is about 13.23 km. The other case optimized the energy consumption over one E&EC drive cycle with and without considering the OBD. The energy consumption from the OBD is assumed to be 0.00497 kWh per OBD event which is equivalent to the energy consumption by spinning the engine for 10 seconds. The OBD event is performed in every 10-second interval when the gen-set does not operate to generate electricity. By doing so, the energy consumption using CS mode only was estimated.

### 2.5. Energy consumption calculation

Understanding the energy flow and energy consumption in different operating modes is critical for HEVs [2, 4, 16]. SAE J1711 standard was used to determine the energy consumption in both CD and CS modes for each trip [7]. After that, the utility factor described in SAE J2841 was used to combine the two energy consumption values into a single energy consumption number for the trip [18]. The equation used to calculate utility factor-weighted energy consumption is expressed below:

$$EC_{UF\_weighted} = EC_{CD} \cdot UF + EC_{CS} \cdot (1 - UF) \quad (17)$$

where $EC_{UF\_weighted}$ denotes the utility factor-weighted energy consumption, $EC_{CD}$ is the energy consumption in CD mode, and $EC_{CS}$ is the energy consumption in CS mode. A charging efficiency of 83% was used when calculating the UF-weighted AC electric energy consumption in the Results and Discussion section.

### 3. Results and discussion
### 3.1. Drive cycle analysis

A drive cycle analysis was performed to understand the difference in both power and energy between the models and vehicle. A comparison of positive power at the wheels is shown in Fig. 8. It is observed that the peak power from the models matches the peak power from the vehicle. However, some of the vehicle dynamics are not captured by the models. We believe this is primarily due to the simplicity of the model and the measurement accuracy in the vehicle.

The differences can also be found in Table 2, where positive propulsion energy, peak power output, average positive power and percent idle time from the simulation and vehicle are compared. The vehicle consumed 15.85% more energy than the models estimated. The average positive power from the vehicle physical test is also 16.60% higher than the average positive power from the models. Interestingly, the models reported a higher peak power output. Since the goal is to develop an energy management strategy that can be used in the vehicle, using the vehicle test data in energy management optimization provides a more representative result. That is why the physical test data was used in the optimization study.

Table 2. Drive cycle results at the wheels

| Test mass: 2800 kg | Unit | Simulation | Phys. Test |
|---|---|---|---|
| Positive propulsion energy | Wh/km | 223.75 | 259.21 |
| Peak power output | kW | 112.50 | 96.96 |
| Average positive power | kW | 14.04 | 16.37 |
| Percent idle time | % | 10.55 | 10.55 |

### 3.2. Energy consumption

The physical test results from the vehicle with the rule-based strategy in three E&EC drive cycles are illustrated in Fig. 9. The first optimization case for the CS mode is also plotted in Fig. 9. The subplot on the top depicts the vehicle velocity. The red and blue curves on the bottom represent the SOC throughout the trip with the rule-based strategy and optimization-based strategy, respectively. The $SOC_{high}$ and $SOC_{low}$ that form the SOC CS window are depicted by the magenta and cyan lines on the bottom subplot, correspondingly. The $SOC_{high}$ and $SOC_{low}$ are 17% and 12%, respectively. The vehicle drove 54.37 km in CD mode. Once the SOC fell below 14.00% SOC, the CS mode kicked in. With the rule-based strategy, the gen-set increased the SOC to 16.94% at the end of the third E&EC drive cycle. The SOC curve from the optimization-based strategy is like the SOC curve from the rule-based strategy. The only difference is the SOC ended at 14.00% at the end of the trip. The rule-based strategy ran the gen-set to put an additional 3.00% SOC into the battery pack, which resulted in a higher energy consumption.

The UF-weighted energy consumption values are listed in Table 3. Both strategies have the exact same energy consumption in CD mode. It is noted that the UF-weighted AC energy consumption includes an 83% charging efficiency from AC to DC. Unlike the UF-weighted energy consumption in CD mode, the optimization-based strategy has a lower UF-weighted fuel energy consumption than the rule-based strategy does in CS mode. There are two major contributors for the difference. The rule-based strategy only operates at a single operating point, while the optimization-based strategy operates at seven operating points and some operating points have a slightly higher efficiency than the operating point used in the rule-based strategy. The other contributor is due to the additional 3% that the rule-based strategy put into the battery pack. The rule-based strategy can be improved by addressing these two areas. Therefore, the rule-based strategy has higher energy consumption values for UF-weighted fuel energy consumption and UF-weighted total energy consumption.

Table 3. CS optimization

|  | Unit | Rule-Based | Optimization-Based |
|---|---|---|---|
| UF-weighted fuel energy consumption | Wh/km | 274.90 | 274.90 |
| UF-weighted AC electric energy consumption | Wh/km | 194.30 | 160.59 |
| UF-weighted total energy consumption | Wh/km | 469.20 | 435.49 |

The results from the OBD study, which is also the second optimization case, are illustrated in Fig. 10. A single E&EC drive cycle simulation was performed for two scenarios: optimization-based strategy without using OBD and optimization-based strategy with using OBD. The black curve on the top represents the vehicle velocity. The magenta and cyan lines on the bottom represent $SOC_{high}$ and $SOC_{low}$, respectively. The red curve is the case without using OBD, while the blue curve represents the case with using OBD. It is observed that both curves overlay most of the time. The SOC difference starts at around 200 seconds and becomes greater than 1% at around 580 seconds. The two curves start merging again at around 880 seconds.

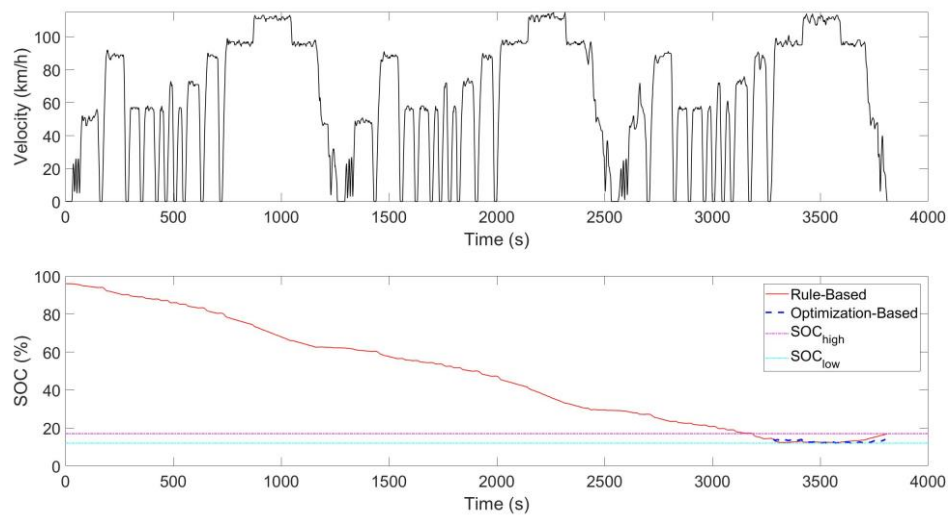

Fig. 9. A comparison of rule-based and optimization-based energy strategies in three laps of E&EC drive cycles

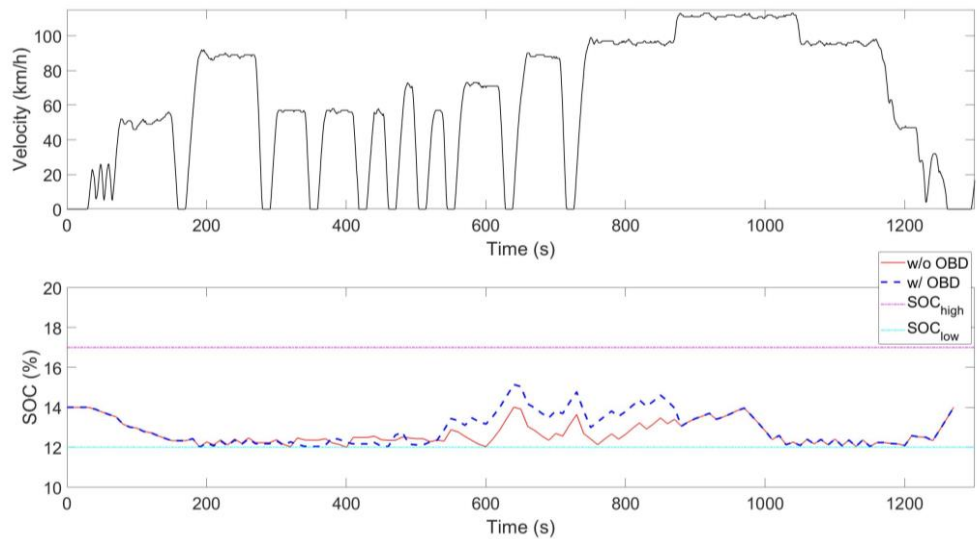

Fig. 10. A comparison of energy management strategy with and without OBD in a single E&EC drive cycle

Table 4 lists the CS energy consumption for both cases. The CS energy consumption with using OBD is slightly higher than the CS energy consumption without using OBD. However, the difference is less than 1.9%.

Table 4. Energy consumption with and without OBD

|  | Unit | w/o OBD | w/ OBD |
|---|---|---|---|
| CS energy consumption | Wh/km | 880.22 | 896.10 |

## 4. Conclusions

The aim of this work was to propose and demonstrate a four-step framework. Unlike previous work, which heavily focused on modeling and simulations, this framework incorporated the utilization of real-world data in the workflow to improve the optimization result. In addition, it identifies areas in the rule-based strategy for future improvement. This work studied the energy management strategy from a different angle and proposed two hypotheses. The first hypothesis is with the measurement errors from in-vehicle measurement system and simplicity of the widely used models, incorporating the physical test data improves the optimization result. The second hypothesis is the extensive OBD events does not increase the energy consumption significantly. To prove the proposed hypotheses, a case study using a PHEV was conducted. The case study showed that the proposed framework provided a more representative optimization result for energy management strategy. In addition, the extensive OBD events only increased 1.9% energy consumption in the E&EC drive cycle.

**Acknowledgements**

This work was partially supported by the US Department of Energy (DOE) Advanced Vehicle Technology Competitions (AVTC).

## Nomenclature

| | |
|---|---|
| AC | alternative current |
| BEV | battery electric vehicle |
| BSFC | brake specific fuel consumption |
| CD | charge depleting |
| CS | charge sustaining |
| DP | dynamic programming |
| E&EC | emission and energy consumption |
| ECMS | equivalent consumption minimization strategy |
| EREV | extended-range electric vehicle |
| EV | electric vehicle |
| FTP | federal test procedure |
| HEV | hybrid electric vehicle |
| HIL | hardware-in-the-loop |
| HV | high voltage |
| HWFET | highway fuel economy driving schedule |
| LHV | lower heating value |
| OBD | on-board diagnostics |
| PHEV | plug-in hybrid electric vehicle |
| SOC | state of charge |

*Optimization of rule-based energy management strategies for hybrid vehicles using dynamic programming*

Di Zhu, DEng. – FREEDM Systems Center, North Carolina State University, United States.
e-mail: *dzhu2@ncsu.edu*

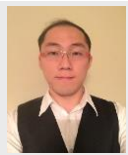

Ewan Pritchard, DEng. – FREEDM Systems Center, North Carolina State University, United States.
e-mail: *egpritch@ncsu.edu*

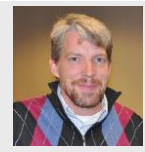

Sumanth Reddy Dadam, MSci. – Mechanical Engineering Department, Wayne State university, United States.
e-mail: *sumanth.me03@gmail.com*

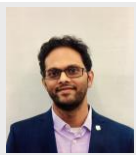

Vivek Kumar, MSci. – Mechanical Engineering Department, Wayne State University, United States.
e-mail: *mechviku@gmail.com*

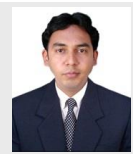

Yang Xu, DEng. – FREEDM Systems Center, North Carolina State University, United States.
e-mail: *yxu17@ncsu.edu*

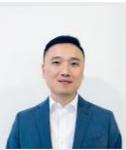